\documentclass[conference,9pt]{IEEEtran}
\IEEEoverridecommandlockouts
\usepackage{multirow}
\usepackage{booktabs}
\usepackage{graphicx}
\usepackage{cite}
\usepackage{amsmath,amssymb,amsfonts}
\usepackage{algorithmic}
\usepackage{graphicx}
\usepackage{textcomp}
\usepackage{xcolor}
\usepackage{hyperref}
\def\BibTeX{{\rm B\kern-.05em{\sc i\kern-.025em b}\kern-.08em
    T\kern-.1667em\lower.7ex\hbox{E}\kern-.125emX}}

\begin{document}
\title{DOSE: Drum One-Shot Extraction from Music Mixture
\thanks{This work was partly supported by the National Research Foundation of Korea (NRF) grant funded by the Korea government (MSIT) [No. RS-2023-00219429, 50\%], Institute of Information \& communications Technology Planning \& Evaluation (IITP) grant funded by the Korea government(MSIT) [No. RS-2022-II220320, 2022-0-00320, Artificial intelligence research about cross-modal dialogue modeling for one-on-one multi-modal interactions, 40\%], [No. RS-2021-II212068, Artificial Intelligence Innovation Hub (Artificial Intelligence Institute, Seoul National University), 5\%], and [NO.RS-2021-II211343, Artificial Intelligence Graduate School Program (Seoul National University), 5\%]
}
}


\author{
\IEEEauthorblockN{
Suntae Hwang$^{1}$ \qquad Seonghyeon Kang$^{1}$ \qquad Kyungsu Kim$^{1}$ \qquad Semin Ahn$^{2}$ \qquad Kyogu Lee$^{1,3}$
}\\[0.5cm]
\IEEEauthorblockA{
$^{1}$ Music and Audio Research Group (MARG), Department of Intelligence and Information, Seoul National University\\
$^{2}$ Department of Mechanical Engineering, Seoul National University\\
$^{3}$ Interdisciplinary Program in Artificial Intelligence \& Artificial Intelligence Institute, Seoul National University\\
\{iamsuntae1, shkang17, kyungsu.kim, susemi2399, kglee\}@snu.ac.kr
}
}

\maketitle
\begin{abstract}

Drum one-shot samples are crucial for music production, particularly in sound design and electronic music. This paper introduces Drum One-Shot Extraction, a task in which the goal is to extract drum one-shots that are present in the music mixture. To facilitate this, we propose the Random Mixture One-shot Dataset (RMOD), comprising large-scale, randomly arranged music mixtures paired with corresponding drum one-shot samples. Our proposed model, Drum One-Shot Extractor (DOSE), leverages neural audio codec language models for end-to-end extraction, bypassing traditional source separation steps. Additionally, we introduce a novel onset loss, designed to encourage accurate prediction of the initial transient of drum one-shots, which is essential for capturing timbral characteristics. We compare this approach against a source separation-based extraction method as a baseline. The results, evaluated using Fréchet Audio Distance (FAD) and Multi-Scale Spectral loss (MSS), demonstrate that DOSE, enhanced with onset loss, outperforms the baseline, providing more accurate and higher-quality drum one-shots from music mixtures. The code, model checkpoint, and audio examples are available at \hyperlink{https://github.com/HSUNEH/DOSE}{\texttt{https://github.com/HSUNEH/DOSE}}

\end{abstract}

\begin{IEEEkeywords}
Drum, One-Shot, Music Source Separation, Neural Audio-Codec, Generative Model
\end{IEEEkeywords}

\section{Introduction}

Drum sounds are fundamental components of contemporary music production, playing a crucial role across various genres such as electronic music, hip-hop, and pop. 
In these genres, music producers often construct rhythmic patterns by sequencing individual drum one-shot samples, allowing for precise control over the timbral and temporal characteristics of the rhythm. 
Given the significance of drum sounds in music production, there has been substantial research interest in synthesizing drum one-shot samples using advanced technologies, including recent developments in deep learning~\cite{neuraldrummachine, neurodrum, drysdale2020adversarial, drumgan, StyleWaveGAN}.
These novel approaches utilize conditional inputs such as drum type or acoustic features, aiming to facilitate intuitive sample creation for music producers without requiring extensive knowledge of signal processing techniques.

In many practical scenarios, music producers work with existing recordings to create remixes, cover versions, or other productions, necessitating the extraction of high-quality drum samples directly from music mixtures.
We define this task as \textit{Drum One-Shot Extraction}.
A conventional approach to this task involves applying music source separation techniques to isolate the drum track, followed by the identification and extraction of segments containing isolated one-shot samples.
We refer to this type of method as the “separation-based approach”. However, this approach cannot guarantee the identification of isolated one-shot segments and is dependent on separation algorithms, potentially introducing artifacts and compromising sample quality.

\begin{figure}[t]
    \centering
    \includegraphics[width=0.9\linewidth]{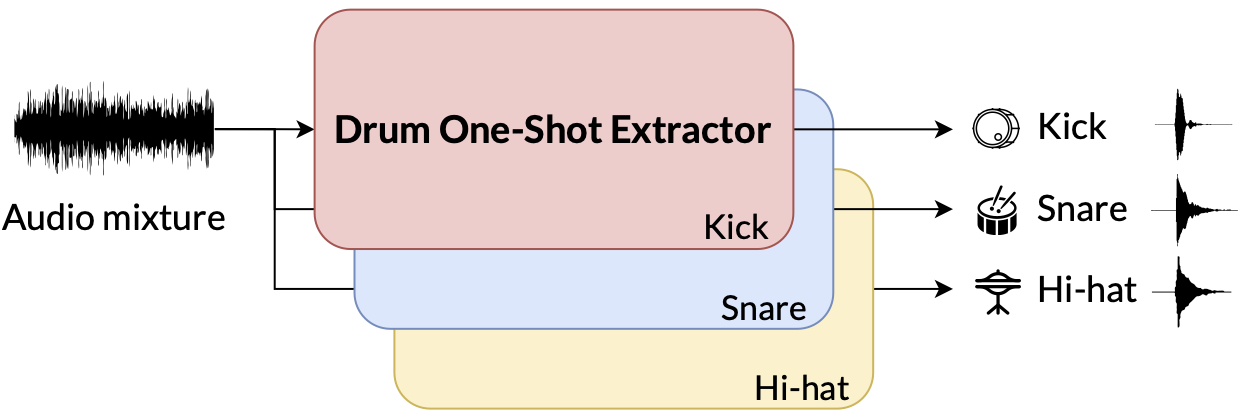}
    \caption{Illustration of our approach. Given an audio mixture as input, each \textit{Drum One-Shot Extractor(DOSE)} model extract one-shot audio samples for kick, snare, and hi-hat drums.}
    \label{fig:overview}
\end{figure}

To address these limitations, we propose a novel \emph{generation-based one-shot extraction} approach that circumvents the intermediate separation step and directly generates drum one-shot samples from the input music mixture. Our method, \textbf{DOSE}, leverages recent advancements in neural audio codec language modeling to perform end-to-end generation of drum one-shots. DOSE employs separate decoder-only Transformers for each drum type (kick, snare, and hi-hat) to achieve better extraction performance compared to using a single model for all drum types.

The architecture of DOSE closely follows that of MusicGen~\cite{copet2024simplecontrollablemusicgeneration} , utilizing the same core components of neural audio codec and decoder-only transformer.
The DAC~\cite{kumarHighFidelityAudioCompression2023} encoder encodes both the music mixture and the one-shot audio into discrete acoustic tokens. 
The transformer is then trained to autoregressively generate the acoustic tokens of the one-shot samples conditioned on the acoustic tokens of the music mixture.
The generated tokens are decoded into waveform audio via the DAC decoder.

For training and evaluation of DOSE, we introduce the Random Mixture One-shot Dataset (RMOD), a novel paired dataset comprising synthetically generated music mixtures and their corresponding drum one-shot samples.
RMOD consists of 360,000 pairs of mixture audio and corresponding drum one-shot samples, created by randomly mixing drum tracks, which are synthesized using one-shot samples, with instrument tracks.

We conducted a comprehensive quantitative evaluation of DOSE and baseline methods on the one-shot drum sample extraction task.
In addition to RMOD, we utilized the Groove MIDI Dataset~\cite{groovemidi} in our evaluation, which offers more realistic drum performances.Our experimental results demonstrate that DOSE outperforms the baseline method, which is a separation-based approach implemented using LarsNet[citation], across various objective metrics, including Fréchet Audio Distance (FAD)~\cite{fad} and Multi-Scale Spectral Similarity (MSS))~\cite{mss}.

\section{Related Work}

\noindent \textbf{Drum One-Shot Generation.}
Generating drum one-shot samples has received considerable attention with the development of neural audio synthesis methods. Early approaches leveraged models such as Variational Autoencoders (VAE)~\cite{Kingma2014} and Generative Adversarial Networks (GAN)~\cite{goodfellow2020generative}, which introduced latent-space exploration and controllable synthesis for drum sounds. Subsequent works like NeuroDrum~\cite{neurodrum}, DrumGAN~\cite{drumgan}, and StyleWaveGAN~\cite{StyleWaveGAN} further improved controllability and timbral diversity by conditioning on various audio features (e.g., brightness, boominess, depth).

Score-based diffusion models~\cite{song2021scorebasedgenerativemodelingstochastic} have recently emerged as a powerful paradigm for audio synthesis, offering flexible sampling strategies. CRASH~\cite{rouardcrash}, for instance, leverages stochastic differential equations (SDEs) to generate high-resolution percussive sounds (44.1\,kHz) in a controllable manner, matching the fidelity of GAN-based methods while enabling techniques such as class mixing to create “hybrid” sounds. These advancements illustrate the growing breadth of approaches for drum/percussive sound generation, providing more interactive and fine-grained creative possibilities for music producers.

\medskip

\noindent \textbf{Drum Source Separation.}
In traditional music source separation, the goal is to split a mixture into individual instrument stems, including drums, bass, vocals, and others~\cite{stoter2019open,hennequin2020spleeter,défossez2019demucsdeepextractormusic,8707065}. More specific to drum-oriented tasks, Mezza et al.\ \cite{Mezza_2024} introduced a new challenge called Drum Source Separation, aiming to decompose a drum track into separate stems (kick, snare, hi-hat, etc). Their proposed method, LarsNet, accomplishes drum-component separation, allowing for individual extraction of each drum type.

\medskip

\noindent \textbf{Neural Audio Codec Language Modeling.}
Neural audio codecs such as SoundStream~\cite{zeghidourSoundStreamEndtoEndNeural2021}, EnCodec~\cite{defossezHighFidelityNeural2022}, and DAC~\cite{kumarHighFidelityAudioCompression2023} have received notable attention for their ability to compress audio into discrete tokens with minimal perceptual loss. When combined with transformer models, these tokenizers can be leveraged for autoregressive audio generation tasks. MusicGen~\cite{copet2024simplecontrollablemusicgeneration} and MusicLM~\cite{agostinelliMusicLMGeneratingMusic2023} are prime examples, demonstrating high-quality music generation via codec-based language models. 

\section{Method}

We propose DOSE, a deep learning model designed to extract drum one-shot sounds (kick, snare, and hi-hat) from complex audio mixtures. 
Inspired by MusicGen~\cite{copet2024simplecontrollablemusicgeneration}, DOSE employs a decoder-only transformer to process discrete audio representations. 
To emphasize accurate transient predictions, we introduce a novel onset loss during training.

\begin{figure}[t]
    \centering
    \includegraphics[width=\linewidth]{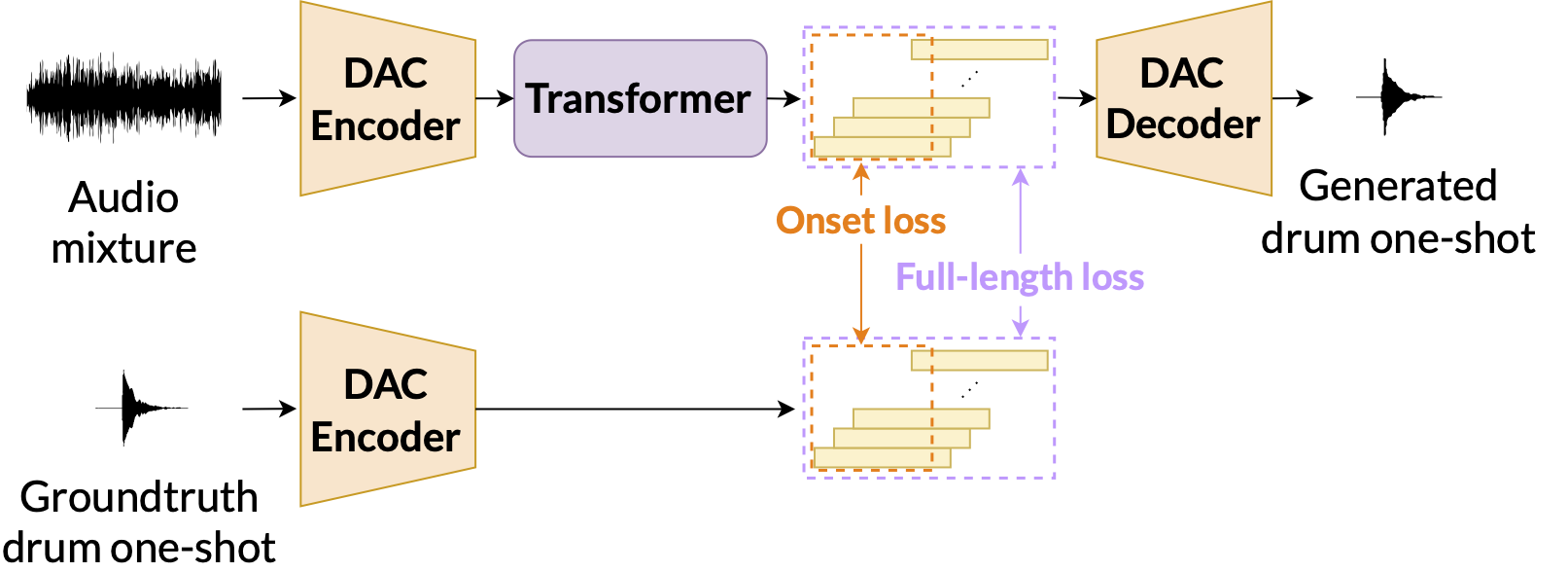}
    \caption{Proposed Method. The input audio mixture is encoded into a sequence of discrete tokens using a frozen DAC encoder, which are then fed into a decoder-only transformer. The transformer is trained to autoregressively predict the groundtruth drum one-shot tokens by minimizing two losses: onset loss and full-length loss. Finally, the predicted token sequence is decoded into drum one-shot audio using the DAC decoder.}
    \label{fig:method}
\end{figure}

\subsection{Autoregressive Acoustic Token Generation}

DOSE generates drum one-shot sounds by autoregressively predicting acoustic tokens conditioned on mixture-audio tokens. 
This involves converting audio waveforms into discrete tokens using Descript Audio Codec (DAC)~\cite{kumarHighFidelityAudioCompression2023}, predicting drum one-shot tokens, and decoding them into waveforms.

DAC processes mono audio input \(x \in \mathbb{R}^{T_{\text{in}} \cdot f_s}\), tokenizing it into a discrete code 
\[
q \in \{1,\ldots,N\}^{(T_{\text{in}} \cdot f_c)\times K},
\]
where \(K\) is the number of codebooks, \(N\) is the codebook size, and \(f_c \ll f_s\) is the codec frame rate.


The decoder-only transformer autoregressively predicts drum one-shot tokens 
\[
\hat{q} \in \{1,\ldots,N\}^{(T_{\text{out}} \cdot f_c)\times K},
\]
conditioned on the mixture-audio tokens. 
The DAC decoder then converts these tokens back to audio waveforms. 
DOSE is trained separately for each drum type (kick, snare, and hi-hat), following the structure of decoder-only language models~\cite{copet2024simplecontrollablemusicgeneration} and employing the interleaving delay pattern from~\cite{kharitonov2022textfreeprosodyawaregenerativespoken}.

\subsection{Training Loss}

Our training objective combines two cross-entropy losses: (1) \emph{full-length cross-entropy}, computed over all predicted tokens in the drum one-shot, and (2) \emph{onset loss}, emphasizing the attack or transient portion by focusing on tokens with indices \(t + k \le K + 1\), where \(t\) is the codec frame index and \(k\) is the codebook index. These transient regions strongly influence perceived timbre~\cite{attack}.

The final loss used for training is summation of the full-length loss and the onset loss:
\[
\mathcal{L} = \mathcal{L}_{\text{full-length}} + \mathcal{L}_{\text{onset}}.
\]
This setup biases the model to focus on transient regions, improving the perceptual fidelity of generated one-shot samples.

\begin{figure*}
    \centering
    \includegraphics[width=0.8\linewidth]{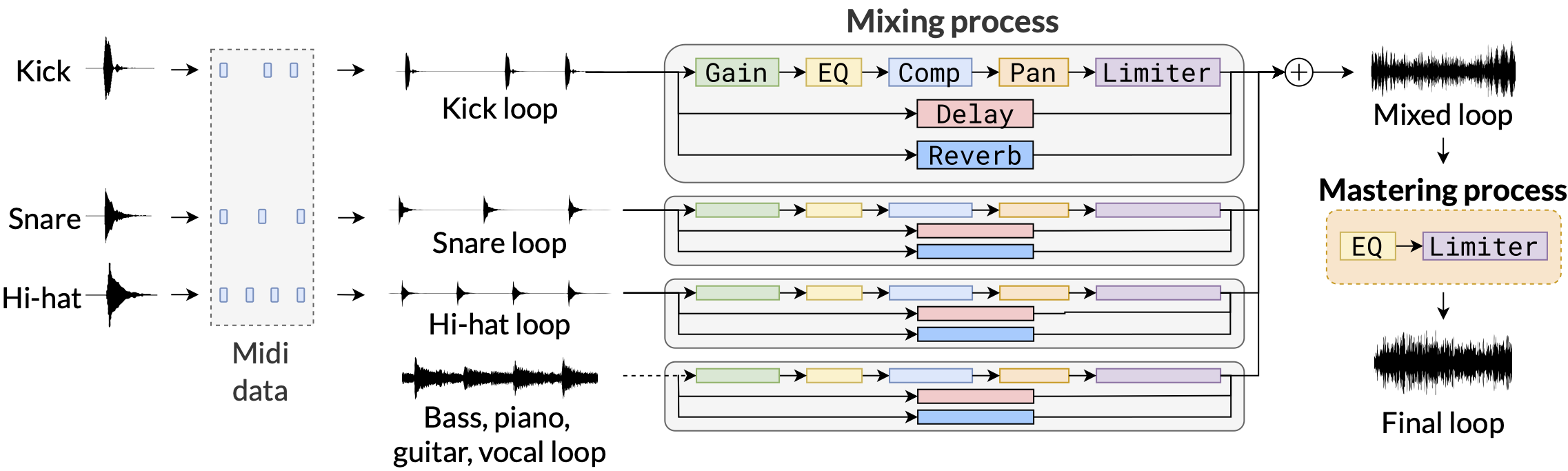}
    \caption{Dataset generation process. First, kick, snare, and hi-hat loops are synthesized from one-shot drum audio samples using randomly generated MIDI notes. Next, optional bass, piano, guitar, and vocal loops are selected. The drum loops and other musical loops are then processed through independent mixing chains, which apply gain, EQ, compression, panning, limiting, delay, and reverb effects. Finally, all tracks are combined and passed through a mastering chain consisting of EQ and limiter effects.}
    \label{fig:datagen}
\end{figure*}

\section{Dataset}

Existing drum sound datasets are not well-suited for one-shot drum extraction, as they often lack proper alignment between mixed audio tracks and their corresponding drum one-shot samples (e.g., kick, snare, hi-hat). This limitation makes it challenging for models to learn one-shot drum extraction compared to drum stem separation tasks, for which more established datasets exist.

To address this gap, we developed the Random Mixture One-shot Dataset (RMOD), a large-scale dataset that includes numerous pairs of randomly mixed music loops and the corresponding drum one-shot samples. These pairs serve as the basis for training and evaluating one-shot drum extraction models.

The overall data generation process for RMOD is illustrated in Figure~\ref{fig:datagen}. Drum one-shot samples were first used to create drum loops, which were then mixed with instrumental loops such as guitar, piano, and bass to form complete musical mixtures.

For the drum one-shot samples, we leveraged publicly available datasets~\cite{drum_and_percussion_kits,one-shot-hip-hop-drums,freesound_one-shot_percussive_sounds}. Instrumental loops were sourced from the Logic Pro~\cite{logicpro2024} library. In total, we collected 3,375 kick samples, 1,801 snare samples, 1,278 hi-hat samples, 454 piano samples, 1,161 guitar samples, 1,782 bass samples, and 202 vocal samples. Using augmentation and mixing techniques described in next sections, we generated one million pairs of randomly mixed music mixtures and their corresponding drum one-shot samples for training, with an additional 10,000 pairs each for validation and testing.

The RMOD dataset has been made publicly available through a dedicated repository, along with detailed documentation to facilitate its use by the broader research community.

\begin{table*}[htbp]
\centering
\caption{Performance Comparison of Models on RMOD and Groove MIDI Datasets using MSS, FAD\_vgg, and FAD\_clap Metrics.}

\begin{tabular}{llccc|ccc|ccc}
\toprule
\textbf{Dataset} & \textbf{Model} & \multicolumn{3}{c|}{\textbf{MSS $(\downarrow)$}} & \multicolumn{3}{c|}{\textbf{FAD\_vgg $(\downarrow)$}} & \multicolumn{3}{c}{\textbf{FAD\_clap $(\downarrow)$}} \\
\cmidrule(lr){3-5} \cmidrule(lr){6-8} \cmidrule(lr){9-11}
 & & \textbf{kick} & \textbf{snare} & \textbf{hihat} & \textbf{kick} & \textbf{snare} & \textbf{hihat} & \textbf{kick} & \textbf{snare} & \textbf{hihat} \\
\midrule
\multirow{4}{*}{RMOD} 
 & Reconstruction & 1.832 & 1.880 & 1.530 & 0.207 & 0.360 & 0.621 & 0.133 & 0.131 & 0.135 \\
 & LarsNet & 5.153 & 6.191 & 5.840 & 4.414 & 6.641 & 3.004 & 1.119 & 1.361 & 1.158 \\
 & DOSE (w/o onset loss) & 4.435 & 4.262 & 3.767 & 1.153 & 1.566 & 1.105 & 0.487 & 0.718 & 0.470 \\
 & DOSE & \textbf{3.700} & \textbf{4.135} & \textbf{2.056} & \textbf{0.920} & \textbf{0.671} & \textbf{0.238} & \textbf{0.439} & \textbf{0.636} & \textbf{0.324} \\
\midrule
\multirow{4}{*}{RMOD (drum only)} 
 & Reconstruction & 1.832 & 1.880 & 1.530 & 0.207 & 0.360 & 0.621 & 0.133 & 0.131 & 0.135 \\
 & LarsNet & 4.841 & 5.932 & 5.695 & 2.738 & 5.553 & 2.503 & 0.887 & 1.207 & 1.028 \\
 & DOSE (w/o onset loss) & 4.518 & 4.316 & 3.793 & 2.409 & 1.649 & 0.937 & 0.497 & 0.706 & 0.448 \\
 & DOSE & \textbf{3.708} & \textbf{3.966} & \textbf{1.972} & \textbf{0.978} & \textbf{0.705} & \textbf{0.217} & \textbf{0.435} & \textbf{0.605} & \textbf{0.301} \\
\midrule
\multirow{4}{*}{Groove MIDI Dataset} 
 & Reconstruction & 2.243 & 1.461 & 1.476 & 2.364 & 3.276 & 1.935 & 0.236 & 0.306 & 0.234 \\
 & LarsNet & 5.079 & 4.411 & 4.752 & 4.674 & 3.286 & \textbf{2.467} & 1.259 & 1.098 & 1.309 \\
 & DOSE (w/o onset loss) & 3.606 & 4.041 &	\textbf{3.356} &	3.453	& 3.799 &	4.491 & 0.690	&1.170	& \textbf{0.624} \\
 & DOSE & \textbf{3.347} & \textbf{3.984} & 3.631 & \textbf{3.369} & \textbf{1.819} & 5.152 & \textbf{0.655} & \textbf{1.089} & 0.933 \\
\bottomrule
\end{tabular}
\label{tab:results}
\end{table*}

\subsection{Drum One-Shot Augmentation}

To increase the diversity of the RMOD dataset and reflect real-world production techniques, we applied drum layering as a data augmentation method.Two drum one-shot samples were randomly selected and layered with amplitude weights of (0.8, 0.2), (0.7, 0.3), and (0.6, 0.4), corresponding to decibel reductions of approximately (-1.94 dB, -13.98 dB), (-3.01 dB, -10.46 dB), and (-4.44 dB, -7.96 dB). This weighted sum approach generated diverse drum combinations. This method effectively expanded the dataset, providing the model with a richer and more diverse set of training examples.

\subsection{Loop Generation Process}

To efficiently generate a large and diverse dataset, RMOD does not aim to closely mimic the structure of real music tracks. Instead, we employed a process of random mixing to create four-second loops, which allows for the rapid creation of a large number of training samples. \\
\textbf{MIDI Generation:} We began by creating MIDI (Musical Instrument Digital Interface) note sequences to control the timing and placement of drum sounds within each loop. 
Using \texttt{miditoolkit}~\cite{miditoolkit} Python library, each four-second loop was divided into 1,920 equally spaced grid points, with kick drum sounds randomly occurring 2 to 4 times, snare drums 2 to 4 times, and hi-hats 14 to 18 times. This random variability in timing and placement provided the model with a wide range of drum sequences, improving its generalization to real-world scenarios. \\
\noindent\textbf{Audio Rendering from MIDI:}
Once the MIDI sequences were created, each note was mapped to a corresponding drum one-shot sample from RMOD. If a new onset occurred for the same drum class while a previous sample was still playing, the earlier sound was sliced to prevent overlap within that class. This behavior reflects real-world scenarios, such as rapid hi-hat strikes, and does not affect overlapping sounds from different drum classes.
For instrumental tracks, we used guitar, piano, bass, and vocal loop samples to introduce additional variability.
Each instrument had a 30\% chance of being excluded when creating the mixed loops, allowing for diverse instrument configurations in the dataset.
To introduce further variation, instrumental loops were randomly sliced into four- or two-second segments. Pitch shifting was then applied using \texttt{librosa}~\cite{mcfee2015librosa} Python library, with bass loops shifted between -6 and +2 semitones, and other instruments shifted between -12 and +12 semitones. \\
\noindent\textbf{Mixing and Mastering Simulation:} 

To reduce the domain gap between the dataset and professionally produced music, we applied a series of digital audio effects during the loop generation process.
In the mixing stage, each instrument and drum track was processed with gain adjustments, equalization (EQ), compression, panning, and limiting, with additional effects such as reverb and delay applied using parallel processing chains.
This processing introduced sufficient variability in the audio characteristics, enhancing the dataset’s diversity to improve the model’s ability to generalize to real-world musical contexts.

During the mastering stage, the mixed audio was subjected to final EQ and limiting adjustments, producing variable outputs reflective of diverse audio environments.
The parameters of these effects were randomized to ensure the dataset covered a broad range of production styles, reducing potential domain mismatch during inference.
All audio files were exported in 16-bit, 44.1 kHz stereo WAV format to maintain consistency and compatibility with standard audio processing tools.
All digital audio effects were implemented using the \texttt{pedalboard}~\cite{pedalboard} and \texttt{pymixconsole}~\cite{steinmetz2020mixing} Python libraries.

\section{Experiments}

In this section, we describe the models under comparison and the datasets used for evaluation. We then report experimental results and analyses.

\subsection{Compared Models}
We evaluate the following models:

\begin{itemize}
    \item \textbf{Reconstruction:} 
    This baseline measures how much distortion arises purely from passing a ground-truth one-shot sample through the DAC encoder and decoder. Since the DOSE also employs DAC to generate outputs, this \emph{reconstruction} score serves as an upper bound on the achievable performance of DOSE.

    \item \textbf{LarsNet:} 
    We employ LarsNet~\cite{Mezza_2024}, a drum source-separation model that decomposes the drum mixture into drum stems (kick, snare, hi-hat). Because we have the drum MIDI data, the onset times of individual drum hits are known. After separating the mixture with LarsNet, we align and slice out a single one-shot drum hit. 

    \item \textbf{DOSE:} 
    Our proposed \emph{generation-based} method that bypasses source separation. We also conduct an ablation study by comparing the DOSE model trained \emph{with} and \emph{without} the onset loss, to assess its impact on generating high-fidelity one-shot samples.
\end{itemize}

\subsection{Datasets}
We use three datasets to evaluate the above models:

\begin{itemize}
    \item \textbf{RMOD:} 
    The \emph{Random Mixture One-shot Dataset} introduced in this paper, which includes one million training samples and 10,000 samples each for validation and testing. Each sample is a 4-second mixture along with paired drum one-shots.

    \item \textbf{RMOD Drums-only:} 
    A subset of the RMOD dataset containing only drum tracks. This subset is used to further evaluate how well each model performs when mixtures primarily consist of drums, reducing the interference from other instruments.

    \item \textbf{Groove MIDI Dataset:} 
    A dataset that contains real human-performed drum MIDI files~\cite{groovemidi_b}. To convert these MIDI tracks to audio, we used \emph{Logic Pro} drum kits \emph{not} used in building the RMOD dataset. This ensures more realistic drum performances and out-of-domain generalization testing for each model.
\end{itemize}

\subsection{Metrics}

We use two objective metrics to evaluate the quality of extracted or generated drum one-shot samples:

\noindent \textbf{Fréchet Audio Distance (FAD).}
FAD~\cite{fad} measures the distance between two distributions of audio embeddings. 
In our experiments, we employ two embedding models : VGGish~\cite{hershey2017cnn}, CLAP~\cite{wu2023large}.
Both versions compare embeddings from the generated audio against a reference distribution, derived from ground-truth drum one-shots in the test set. 

\medskip

\noindent \textbf{Multi-Scale Spectral Similarity (MSS).}
MSS~\cite{mss_b} evaluates how closely a generated audio sample matches a reference audio sample in terms of its time-frequency representation. 
To compute the multi-scale spectral similarity (MSS), we first transform the generated and reference signals into spectrograms at multiple scales, specifically using FFT window lengths of 2048, 1024, 512, 256, 128, and 64.
We then compute the mean squared error (MSE) between the spectrograms at each scale and aggregate them. 

\subsection{Results and Discussion}
Table~\ref{tab:results} summarizes the performance of each model across RMOD, RMOD Drums-only, and the Groove MIDI Dataset. We report both MSS and FAD (with VGG and CLAP embeddings). The following key observations emerge:
\begin{itemize}
    \item Reconstruction exhibits a lower bound of the codec’s inherent loss, as it simply measures the quality of passing the ground-truth one-shot through the DAC encoder-decoder.
    \item DOSE significantly outperforms LarsNet, particularly for high-frequency percussion (e.g., hi-hat). The version of DOSE trained with onset loss further improves transient accuracy, reducing spectral errors and improving perceptual quality.
    \item However, DOSE’s performance in extracting hi-hat sounds declines on the Groove MIDI Dataset. This degradation is likely due to the presence of additional percussive elements (e.g., toms, crash cymbals, rims) that are absent in RMOD’s focused kick–snare–hi-hat configuration. These extra percussion sources introduce complex spectral overlaps and transient interferences, making it more challenging for DOSE to accurately extract hi-hat sounds.
\end{itemize}

\section{Conclusion}
In this paper, we introduced the task of Drum One-Shot Extraction, aiming to generate drum one-shot samples from a given music mixture. To tackle this task, we proposed the Random Mixture One-Shot Dataset (RMOD), containing one million training samples and 10,000 validation and test samples, each consisting of a music mixture paired with the corresponding drum one-shot samples.

We also introduced Drum One-Shot Extractor (DOSE), a neural audio codec-based model designed to generate high-quality drum one-shots from complex music inputs. Using objective metrics such as Fréchet Audio Distance (FAD) and Multi-Scale Spectral Loss, DOSE outperformed the baseline model, LarsNet, across multiple datasets, demonstrating its ability to generate perceptually accurate drum sounds.

A limitation of this work is the lack of paired data from real commercial music, which we aim to address in future work. Additionally, we plan to extend this approach to other instruments, enabling the generation of high-quality one-shots for a broader range of instruments, further enhancing music production possibilities.

\newpage

\bibliographystyle{ieeetr}

\end{document}